\documentclass[appendixfloats,tighten,iop,numberedappendix]{emulateapj}

\begin{document}
\title{A Tentative Size-Luminosity Relation for the Iron Emission-Line Region in Quasars}
\author{Doron Chelouche\altaffilmark{1},  Stephen E. Rafter\altaffilmark{1,2},  Gabriel I. Cotlier\altaffilmark{1}, Shai Kaspi\altaffilmark{3,2}, and Aaron J. Barth\altaffilmark{4}}
\altaffiltext{1} {Department of Physics, Faculty of Natural Sciences, University of Haifa, Haifa 31905, Israel; doron@sci.haifa.ac.il}
\altaffiltext{2}{Physics Department, Technion, Haifa 32000, Israel; e-mail: rafter@physics.technion.ac.il}
\altaffiltext{3}{School of Physics \& Astronomy and the Wise Observatory, Tel-Aviv University, Tel-Aviv 69978, Israel; shai@wise.tau.ac.il}
\altaffiltext{4}{Department of Physics and Astronomy, 4129 Frederick Reines Hall, University of California, Irvine, CA 92697, USA; barth@uci.edu}
\shortauthors{Chelouche et al.}
\shorttitle{A Size-Luminosity Relation for the Iron-Emitting Region in Quasars}

\begin{abstract}

New reverberation mapping measurements of the size of the optical iron emission-line region in quasars are provided, and a tentative size-luminosity relation for this component is reported. Combined with lag measurements in low-luminosity sources, the results imply an emission-region size that is comparable to and at most twice that of the H$\beta$ line, and is characterized by a similar luminosity dependence. This suggests that the physics underlying the formation of the optical iron blends in quasars may be similar to that of other broad emission lines. 

\end{abstract}

\keywords{
galaxies: active ---
methods: data analysis ---
quasars: emission lines
}

\section{Introduction}

The geometry of the broad line region (BLR) in quasars can be studied by means of reverberation mapping, where one tracks flux variations in the emission line in response to continuum fluctuations \citep{np97}. The distance range subtended by the BLR gas implies a highly-stratified medium with high-ionization lines, such as \ion{He}{2}\,$\lambda 1640$ and \ion{C}{4}\,$\lambda 1548$, being formed close to the central continuum-emitting source, possibly on scales comparable to the outer optical-emitting accretion disk \citep[and references therein]{rod97,pet99,c13}, and low ionization species, such as \ion{H}{1}, being emitted from larger regions whose size is comparable to the dust sublimation radius \citep{nl93,sug06}. Complicated radiative transfer physics may also affect the apparent size of the BLR \citep{ben10}. Accumulating statistics have shown that the effective area of the BLR scales in proportion to the quasar luminosity \citep{ben09}. This relation has been relatively well established for the Balmer and \ion{C}{4}\,$\lambda 1549$ emission lines \citep{kas00,kas07}, but has not yet been shown to hold in general.

Among the most prominent spectral features in the (rest) optical-UV spectra of quasars are the iron emission blends \citep{bg92}: poorly resolved plethora of numerous emission lines predominantly associated with \ion{Fe}{2}, which can only be partially resolved in narrow line objects \citep{vero04}. Despite several decades of intensive research in the field \citep[and references therein]{wam67,sar68,ok76,col79,net80,gr81,kk81,nw83,wil84,bal04,ver04,zh07,bv08,fer09,sh10,do11}, relatively little is known about the physics of these features. For example, it has been argued that collisional excitation, rather than photo-excitation, is responsible for the bulk of the iron emission \citep{col00}, setting its physics apart from the rest of the BLR \citep[but see][]{ves05}. Further, many models including state-of-the-art atomic data and detailed radiative transfer calculations fall short of explaining the phenomenological properties of those blends. Reverberation mapping of this component has proven difficult \citep{ves05,kue08} with only very recent works being able to place the optically-emitting iron blend region around the Balmer line region in a few objects  \citep{bia10,bar13,raf13}. Interestingly, \citet{hu08} find that the apparent kinematics of the iron blends differs from that of other emission lines, providing interesting clues about the BLR physics \citep[see however \citealt{sul12}]{fer09}. 

Here we report new iron blends' lag measurements for the Palomar-Green sample of quasars from \citet{kas00}, and quantify the size-luminosity relation over four decades in luminosity. Our analysis makes use of (and extends) the multi-variate correlation function (MCF) scheme of \citet{cz13}, and is shown to work in cases where reliable spectral decomposition is difficult to achieve. This paper is organized as follows: the MCF scheme is summarized and extended in section 2. Results for the PG quasars are outlined in section 3, with the discussion following in section 4.

\begin{figure*}
\plottwo{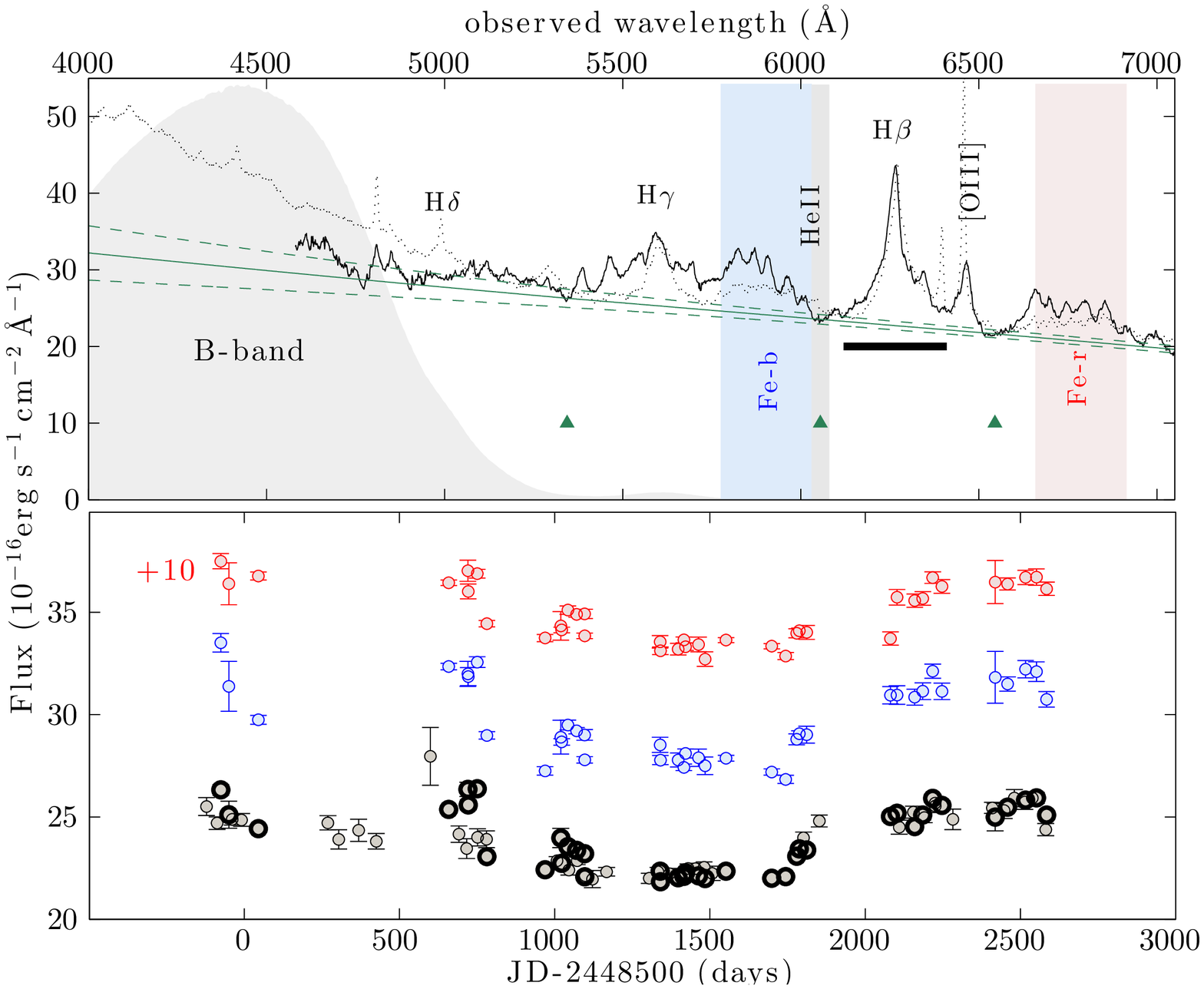}{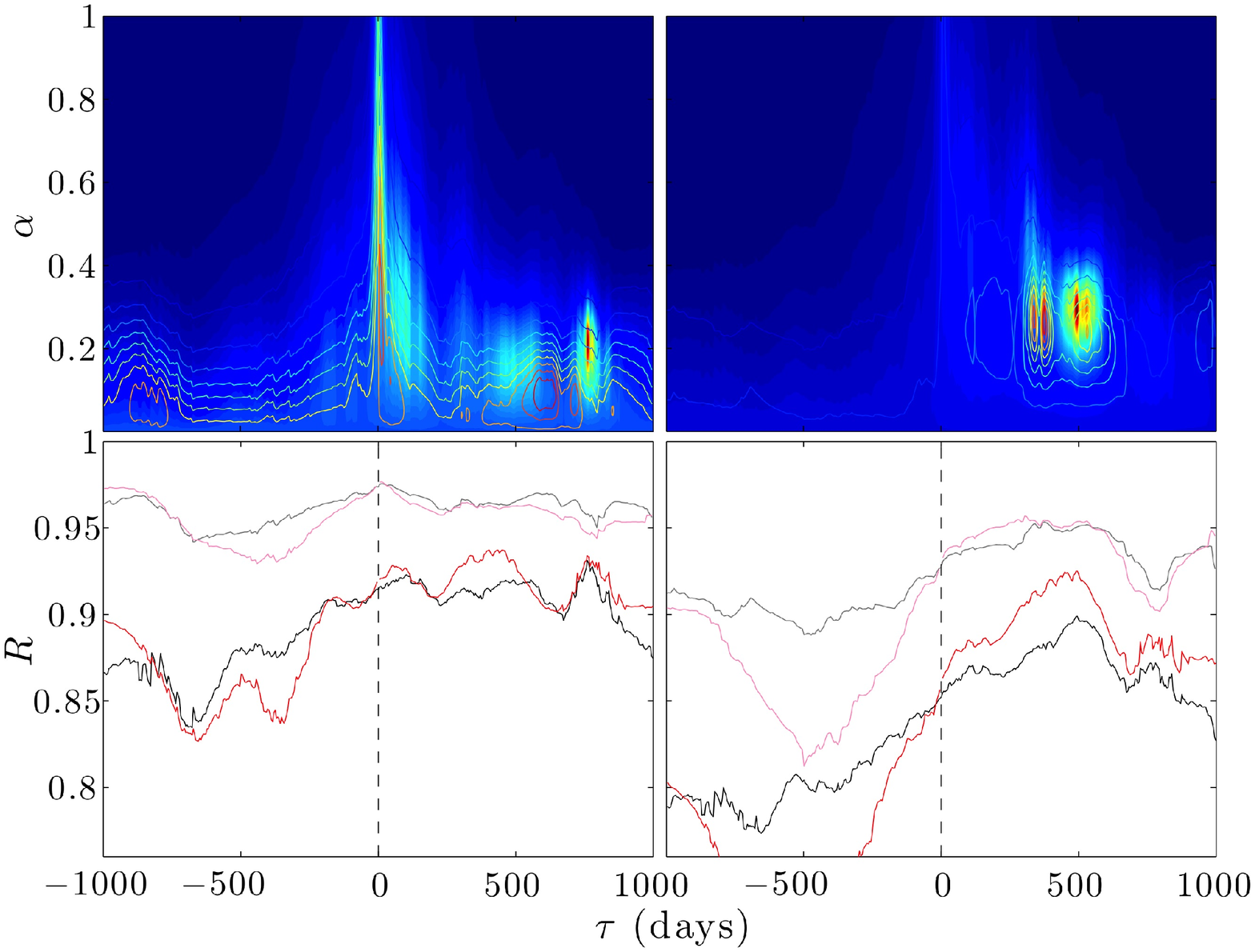}
\caption{Spectrum, bandpasses, light curves (left panels) and correlation analyses (right panels) for PG\,1700+518. The mean spectrum for PG\,1700 from the \citet{kas00} dataset (solid line), with the quasar composite spectrum of \citet{dvb01} over plotted (dashed line) in the top left panel. The $B$-band transmission curve is also shown, as are the bands from which the light curves used in this analysis are derived (gray for $f_c$ or $f_c^{sp}$, and blue and red for either of the iron blends). Green triangles mark line-free regions used to fit the underlying continuum with second-order polynomials (median fit values are shown as a solid green curve with models deviating by one standard-deviation traced by the dashed green lines). The wavelength region from which information about H$\beta$ has been extracted is marked by a thick horizontal black line (c.f. table 2 in \citealt{kas00}). Spectro-photometric light curves are denoted by gray points and are shown in the bottom-left panel with the pure spectroscopic continuum light curve marked in thick black circles. The blue and red iron-blend  light curves are shown in their respective colors, and are arbitrarily shifted for clarity. Two-dimensional (2D) correlation functions are shown in the right panels: left and right columns for the blue and red iron blends, respectively. Filled contours show the correlation function as calculated using the spectrophotometric light curves, with regular contours marking its values as calculated using the spectroscopic data (warmer colors correspond to larger correlation coefficients). Projected correlation functions are shown in the bottom panels with bright/faded colors corresponding to spectrophotometric/spectroscopic data. Red shades show the corresponding analyses with the kernel version of the correlation function (KMCF, whose 2D versions are not shown).}
\label{pg1700}
\end{figure*}

\section{Method}

Given a continuum light curve in some spectral band, $f_c$, and a light curve consisting of the combined contribution of continuum and lines, $f_{cl}$, we construct a model for the latter. The line is assumed to linearly react to $f_c$ (e.g., the case of photo-excited gas around equilibrium) with a line transfer function, $\psi$, which reflects on the gas geometry, and  whose light crossing-time centroid is $\tau$. Our model for $f_{cl}$ is 
\begin{equation}
f_{cl}^m=(1-\alpha)f_c+\alpha f_c*\psi(\tau),
\label{mo}
\end{equation}
where $\alpha$ reflects on the contribution of the emission line to the total flux in the band and the last term denotes convolution. Determining the lag, $\tau$, is then reduced to finding the best match between $f_{cl}^m$ and $f_{cl}$ using, e.g., the MCF formalism. In \citet{cz13} we chose $\psi=\delta (t-\tau)$ so that $f_c*\psi=f_c(t-\tau)$. Here we also consider a model where $\psi$ has finite structure: a rectangular shape in the time range $[0,2\tau]$ \citep{lsst}. While the true $\psi$ is poorly known, and various models exist \citep{wel91}, they all have a common feature: upon convolution, continuum light curve fluctuations on timescales  shorter than $\tau$ are suppressed. This results in a more physically-motivated $f_{cl}^m$, which is in better qualitative agreement with observations \citep[see figure 1 in][where the iron light curves are smoother compared to the continuum light curves for NGC\,4593]{bar13}. In this paper, results using the new formalism, which we term kernel-MCF (KMCF), are used to corroborate the MCF measurements, and the implementation of the algorithm is identical to that which is described in \citet{cz13}.

We adopt a very conservative approach to estimate the significance of our results, mainly due to the small number of visits in the PG light curves. In particular, we have implemented a flux-randomization, random subset selection (FR-RSS) algorithm for the (K)MCF scheme following \citet[see below]{pet98}. Nevertheless, for reasons discussed in \citet{cz13}, we refrain from quoting its measurement uncertainties.

\begin{figure*}
\plottwo{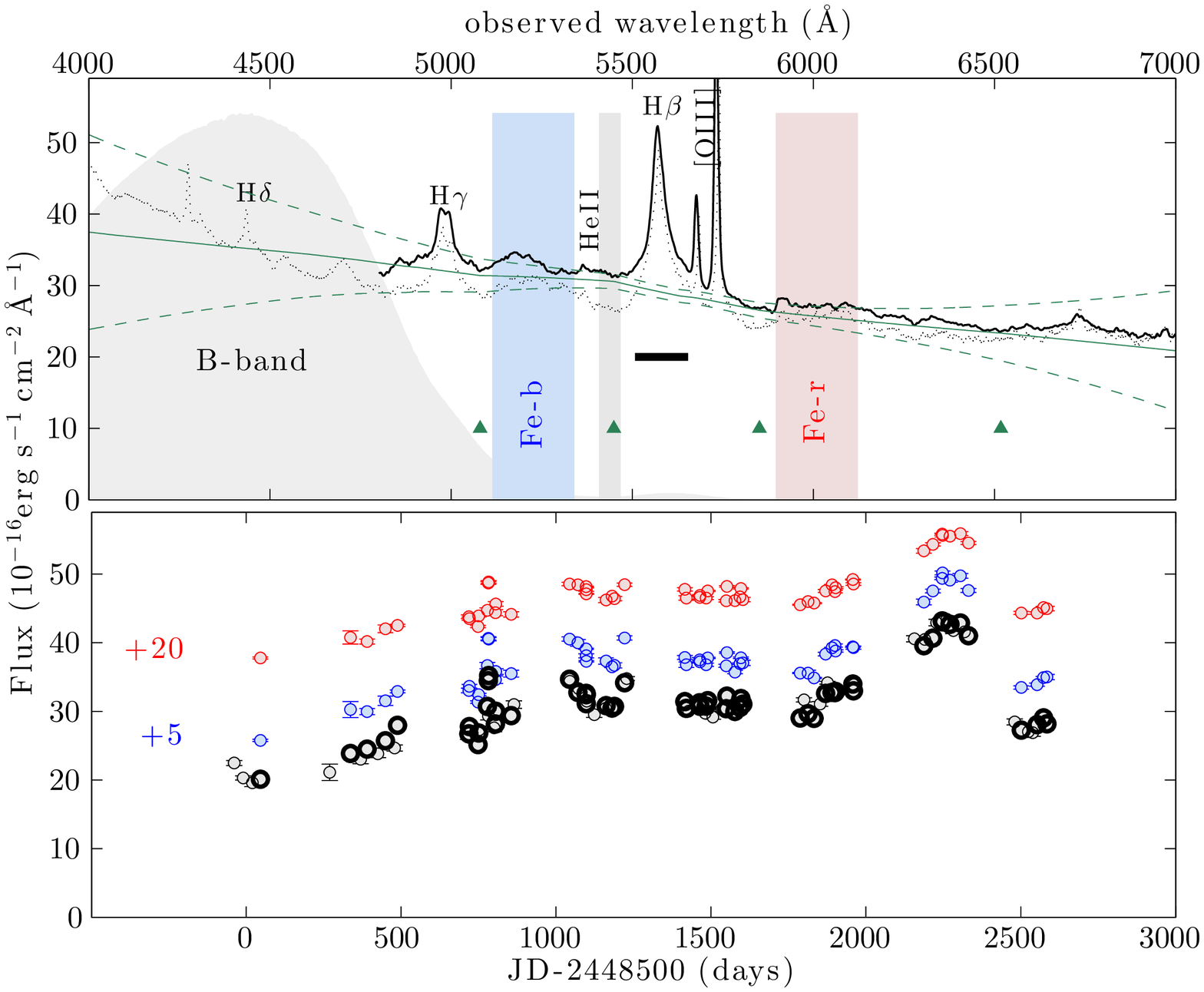}{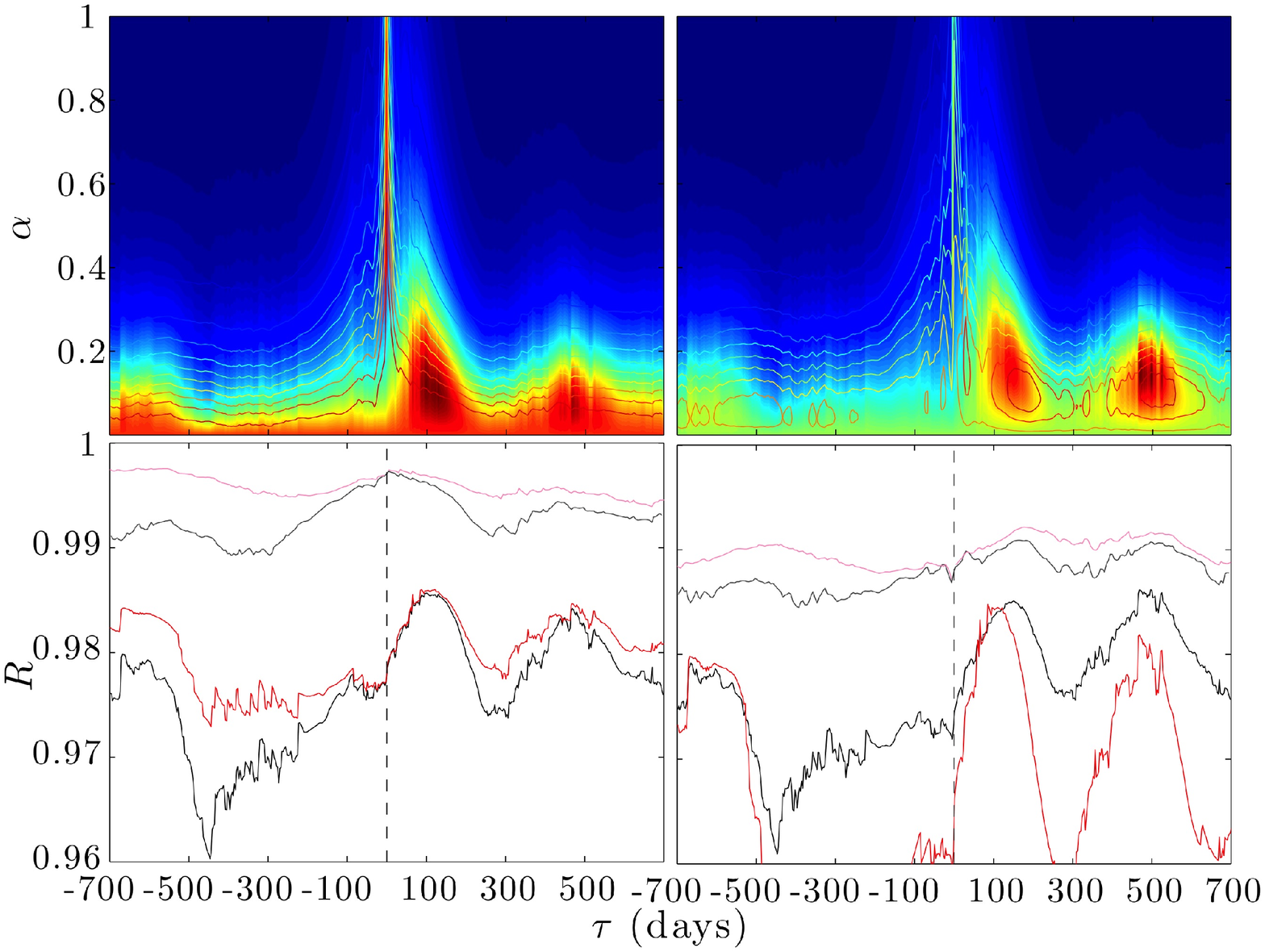}
\caption{Same as figure 1 for PG\,0026+129.}
\label{pg0026}
\end{figure*}

\begin{figure*}
\plottwo{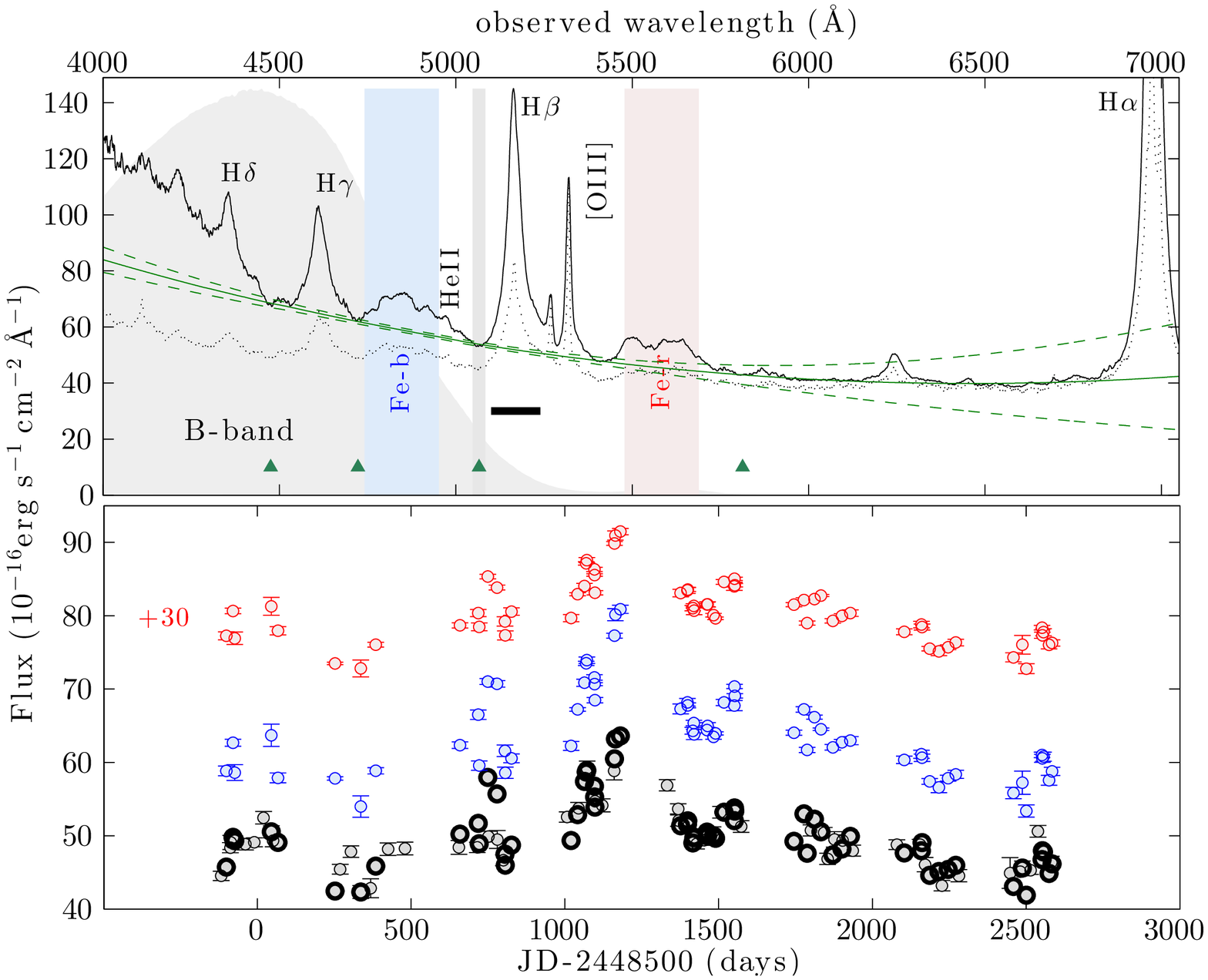}{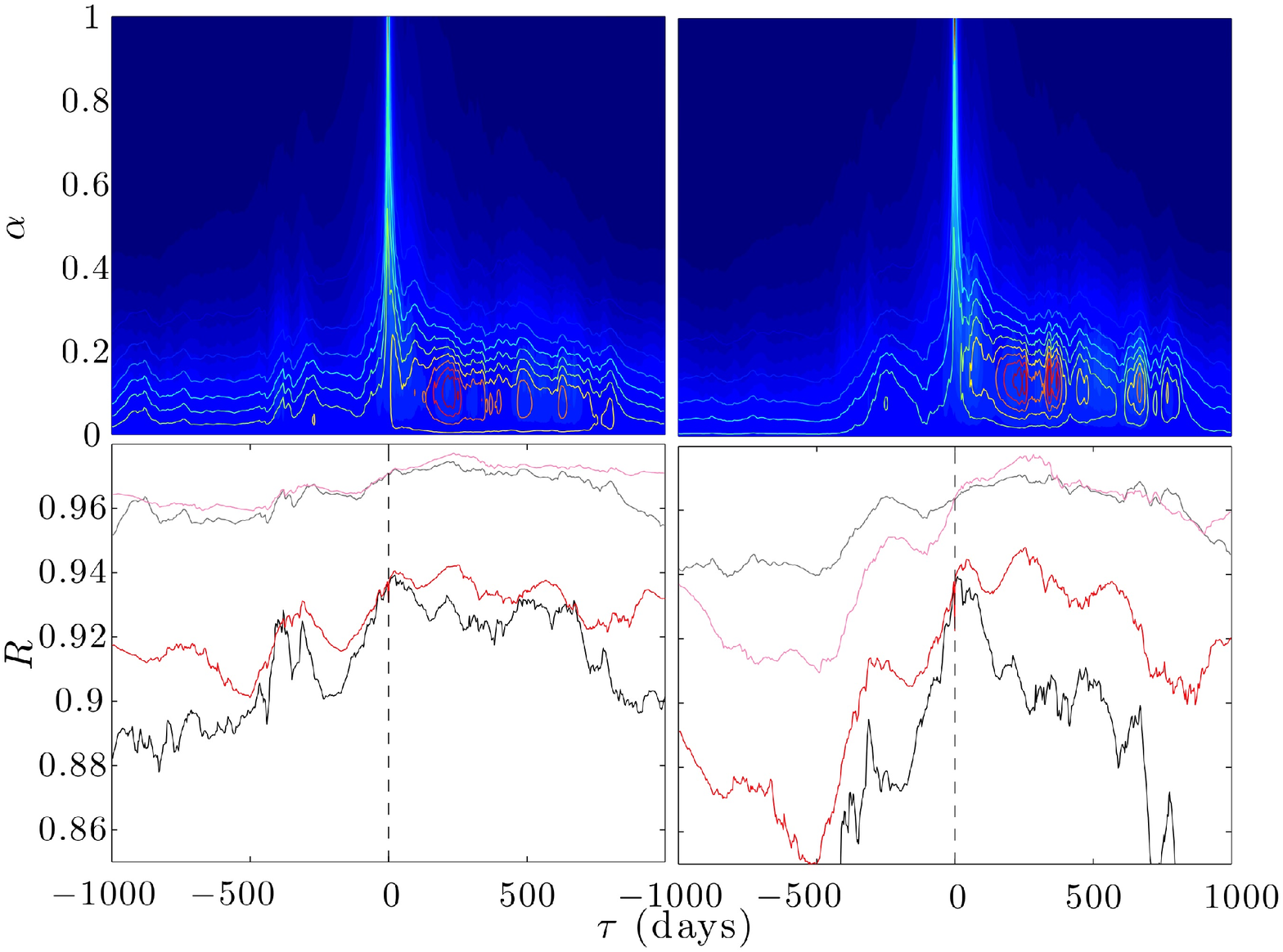}
\caption{Same as figure 1 for PG\,2130+099.}
\label{pg2130}
\end{figure*}

\section{The PG sample of quasars}

We consider the sample of 17 PG quasars from \citet{kas00}, which provided the first reliable Balmer line reverberation mapping results for luminous active galactic nuclei. We use published spectro-photometric light curves that trace the continuum emission over 7.5\,years, and term those $f_c^{sp}$. To mitigate the contribution of emission lines and blends to the $B$-band, hence to $f_{c}^{sp}$, we consider an additional version of the continuum light curves, $f_c^s$, which consists of spectroscopic visits, naturally resulting in fewer points (figure\ref{pg1700}). $f_{cl}^m$ is then separately evaluated using $f_c^s$ and $f_c^{sp}$, and the results compared. 

In addition, we define two iron-rich spectral windows around the H$\beta$ line (see figure \ref{pg1700} for the case of PG\,1700+518, hereafter PG\,1700), and create a set of light curves ($f_{cl}$) in those bands. The light curves trace the mean flux level in the relevant spectral window, and the relative flux error is conservatively matched to that of the continuum light curve. The rest-frame wavelength range for the blue iron blend is 4468\AA-4668\AA\ (i.e., short ward of the \ion{He}{2}\ line), and that for the red iron blend is 5150\AA-5350\AA. As the rest equivalent width of the iron blends is of order $20$\AA\ \citep[see their table 2]{dvb01}, their relative contribution to the flux in our spectral windows, $\alpha^s$, is at the 10\% level. Table 1 lists $\alpha^s$ values for individual objects and blends, as obtained by fitting the underlying continuum by second-degree polynomials (uncertainties were estimated by carrying out a random subset selection algorithm on the number of data points used for fitting purposes; see figure 1).

\subsection{Results}

Among the 17 objects in \citet{kas00} we report here only the three best cases. To be included, at least one of the lag estimates should have an FR-RSS probability of $<15$\% of being spurious (i.e., negative) for either the blue or the red blends, involving either $f_c^s$ or $f_c^{sp}$, and using either the MCF or KMCF schemes. It turns out that only quasars with the best sampling (relative to the time delay) and with a relatively strong iron blends' contribution to the flux make it into the sample: PG\,1700, PG\,0026+129 (PG\,0026), and PG\,2130+099 (PG\,2130). Their results are summarized in table 1 and discussed below. Less robust lag measurements, in the FR-RSS sense, for additional PG objects are further mentioned in section 4.

The analysis of PG\,1700 shows a highly-significant non-ambiguous solution for the red iron blend in the MCF (figure \ref{pg1700}) whether $f_c^s$ or $f_c^{sp}$ are used\footnote{Repeating the analysis for a $f_c^s$ version that was extracted from a wavelength region further removed from the \ion{He}{2}\,$\lambda 4687$ line, led to consistent results.}. While the latter includes a small contribution from high-order Balmer lines and continuum emission that enter the $B$-band, their combined contribution to $f_c^{sp}$ is smaller than the iron's to $f_{cl}$ (see below). This, combined with the expected shorter delays for the higher order Balmer emission \citep{ben10} result in a meaningful lag measurement \citep{lsst}. Reassuringly, the KMCF and MCF give statistically-consistent results suggesting that the signal is not due to short duration features in the light curves. The deduced value of $\alpha$ is on the high-end side, indicating that some 25\% of the (varying) flux in the spectral window is from iron. This is consistent with the mean spectrum of this source showing more prominent iron emission than typical of quasars (Table 1). Averaging the time-lag measurements for the red blend, we find a lag of $\sim 380$\,days, which is $\sim 3$ times longer than the H$\beta$ lag for this source yet statistically consistent with it \citep{kas00}. The fact that this timescale is of order a year appears to be a coincidence since it does not correspond to the seasonal gaps or their multiplicities. Further, the deduced lag of $\sim 380$\,days is consistent with the findings of \citet{bia10} who measured a lag of $270^{+130}_{-190}$\,days (observed frame) using an independent template-fitting scheme. 

\begin{table*}
\tiny
\begin{center}
\caption{Time-lag measurements}
\renewcommand{\tabcolsep}{0.11cm}
\begin{tabular}{lllllllllllll}
\tableline
\rule{0pt}{3ex} & & \multicolumn{3}{c}{Fe-b} & & \multicolumn{3}{c}{Fe-r} \\ \\
\cline{3-5}  \cline{7-9} 
\rule{0pt}{3ex} & $\left < t \right >$ & $\tau$  & & & & $\tau$ & & & $\tau$(Fe) & $\tau$(H$\beta$) \\
Object  & (days) &  (days) & $\alpha$ & $\alpha^s$ & &  (days) & $\alpha$ & $\alpha^s$ & (days) & (days)  \\
(1) & (2) & (3) & (4) & (5) & & (6) &  (7) & (8) & (9) & (10) \\
\tableline

\rule{0pt}{3ex}PG\,0026+129  & 25 & $40^{+240}_{-30}$\,($-630^{+650}_{-70}$) &  $0.05\pm0.07$ & $0.07\pm0.03$ & &$170^{+20}_{-130}$\,($160^{+340}_{-660}$) &  $0.12\pm0.05$  & $0.05\pm0.05$ & $152^{+18}_{-27}$ & $165\pm30$   \\
$[z=0.142\,,L_{45}=0.70]$& & $90^{+50}_{-10}$\,($100\pm50$) &  $0.12\pm0.02$ & & & $145^{+6}_{-10}$\,($125^{+25}_{-15}$) &  $0.17\pm 0.02$  \\
\tableline

PG\,1700+518 &  32 & $0_{-80}^{+300}$ ($0_{-200}^{+500}$) &  $<0.3$ & $0.21\pm0.07$ & & $370_{-70}^{+160}$ ($314_{-140}^{+200}$) &  $0.25\pm0.05$  & $0.19\pm0.03$ & $382^{+76}_{-68}$ &  $250^{+80}_{-100}$ \\
$[z=0.292\,,L_{45}=2.71]$ & & $100_{-100}^{+360}$ ($400_{-340}^{+50}$) &  $<0.3$ & & & $385_{-160}^{+140}$ ($458_{-115}^{+50}$) &  $0.29\pm0.06$  \\ 
\tableline

PG\,2130+099  & 22 & $220_{-20}^{+30}$ ($230\pm 30$) &  $0.1\pm0.02$ & $0.16\pm0.02$ & & $230_{-40}^{+140}$ ($250_{-20}^{+50}$) &  $0.13\pm0.02$ & $0.19\pm0.07$ & $195_{-145}^{+55}$ &  $160^{+150}_{-120}$ \\
$[z=0.061\,,L_{45}=0.22]$ & & $25_{-10}^{+25}$ ($237^{+20}_{-200}$) &  $<0.2$ & & & $50_{-30}^{+150}$ ($250\pm30$) &  $<0.2  $   \\  

\tableline
\end{tabular}
\end{center}
Columns: (1) object properties [name, redshift, and monochromatic optical luminosity $L_{45}=\lambda L_\lambda (5100\,{\rm \AA})/10^{45}\,{\rm erg~s^{-1}}$], (2) the median sampling period of the spectroscopic time series. MCF solutions are shown for the lags (columns  3,6) and the relative flux contributions of the two iron blends (columns 4,7) given the waveband definitions in figures 1-3. Lag measurements in parenthesis correspond to the KMCF analysis results. Spectrally derived relative flux contributions of the two iron blends to the defined wavebands, $\alpha^s$, are shown in columns 5 and 8 with their uncertainties drawn from those of the continuum fits (Figures 1-3). All (K-)MCF uncertainties were deduced using the flux-randomization scheme of \citet[including the 15 to 85 percentiles]{cz13}, but only objects leading to secure lag detections in the FR-RSS sense are reported (see text). The $\alpha$ distributions are Gaussian to a good approximation hence the symmetric uncertainty intervals. Two rows appear per object: the upper/lower reports on the results using the $f_c^s/f_c^{sp}$ light curves. Upper limits on $\alpha$ indicate cases where a prior has been set to constrain the lag (the quoted prior is set per object given the MCF results in the other blend or using the $f_c^s$ light curves; note the agreement with columns 5, 8). (9) Non-weighted mean lag of all the red iron blend measurements with the reported uncertainty range bracketing the extreme values obtained. (10) The H$\beta$ time delays were obtained here, as in column (9), using the same (k-)MCF analyses as for the iron blends (not shown).
\end{table*}

Results are less significant for the blue iron blend in PG\,1700, but are still largely consistent with those obtained for the red one. Specifically, the MCF solutions are multi-peaked (note the different maxima in the 2D plane of figure \ref{pg1700}). Using a spectrally-motivated prior of $\alpha<0.3$, discards the large $\alpha$ and $\tau\sim 0$\,days solution, leading to a more meaningful lag determination, albeit with a large uncertainty using both the MCF and the KMCF schemes.

The case of PG\,0026, shows multi-peak solutions with the first peak corresponding to the delay, and subsequent peaks extending to positive times (not shown in full succession in figure \ref{pg0026}) being spaced by one year intervals, about twice the seasonal gap for this source. Focusing on the first peak for the red iron blend, which is consistent between the MCF and KMCF, we find a delay of $\gtrsim 100$\,days (table 1), i.e., similar to the H$\beta$ delay for this source \citep[table 6 in][]{kas00}. Results for the blue iron blend are less robust and point to somewhat shorter delays (especially when using $f_c^s$), with the KMCF and the MCF providing similar results. The deduced $\alpha$ is consistent with spectroscopic measurements. 

Lastly, we turn to PG\,2130 (Fig. \ref{pg2130}) for which two seemingly contradicting time-delays exist in the literature for the Balmer emission lines \citep{kas00,gr08}. We do not seek to resolve this problem in the present work, but note that we are using the \citet{kas00} dataset and so are sensitive only to long delays.  Results involving spectrophotometric data are less-reliable (with an apparent advantage to the KMCF algorithm; table 1) since the contribution of the iron blend itself (as well as prominent higher order Balmer line emission) to the $B$-band is considerable. For this reason, and given the results using $f_c^s$, which indicate $\alpha\sim 0.1$, we measure the lag using $f_c^{sp}$ by setting the prior $\alpha<0.2$ (see table 1). We find a delay of order 200\,days in most cases (the MCF with $f_c^{sp}$ is an exception, leading to shorter lags by a factor $>4$), which is consistent with the H$\beta$  region size found by \citet[see their table 6]{kas00}.

\section{Discussion and Conclusions}

We reported time-delay measurements for the optical iron blends in three PG quasars. Taken at face value, and noting the slight advantage of the KMCF algorithm in deducing the lag for the blue iron blend, both iron blends appear to originate from regions of comparable sizes. In what follows we define the mean delay for the iron blend region with an uncertainty bracketing the range covered by the various measurements, all of which are assumed to have similar statistical weights (table 1)\footnote{These uncertainties should {\it not} be considered as proper measurement uncertainties.}. Our sample roughly doubles the number of sources with "reliable" lag determinations for the optical iron blends.

\begin{figure}
\plotone{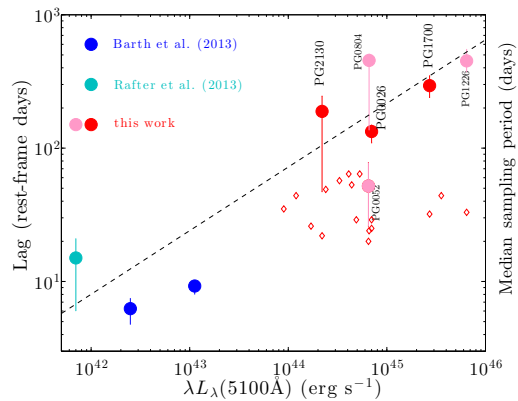}
\caption{The rest-frame lag-luminosity diagram for the optical iron emission blends in active galactic nuclei (see legend).  Dashed line is the  least-squares fit to all reported iron measurements ($K=-19.13$ and $\alpha=0.477$; see text). Light shaded red points correspond to results which are insignificant  according to the FR-RSS scheme. Red diamonds mark the typical sampling period of the spectroscopic time series for all 17 quasars in the \citet{kas00} sample. Error bars on the measurements presented in this work should not be treated as formal measurement uncertainties (see text). }
\label{rl}
\end{figure}

\subsection{A tentative size-luminosity relation}

Figure \ref{rl} shows a size-luminosity diagram for the iron blend in quasars using our results and those reported for the blue iron blend by \citet{bar13}\footnote{The optical luminosity was determined from the continuum model of \citet[see their Fig. 2]{bar13} and using standard $\Lambda$CDM "737" cosmology. We note that \citet{ben13} report a considerably higher luminosity for NGC\,4593 than Barth et al., which could reflect on the source's flux state at that particular epoch.} and \citet{raf13} for low-luminosity objects. If our measurements are in the right ballpark then they imply, for the first time, a size-luminosity relation for the optical iron-emitting region, whose powerlaw index $\sim 0.5$, hence consistent with similar relations for other lines \citep{kas00,kas07}. Specifically, a fit of the form ${\rm log}(R_{\rm BLR})=K+\alpha {\rm log}(\lambda L_\lambda (5100\,{\rm \AA}))$ to the entire data set (see below) yields\footnote{We estimated the uncertainty on the fit parameters using a value randomization (assuming a uniform distribution over the quoted error intervals for each point) random subset selection scheme. The small size of the sample and the uncertainty in the quoted measurement errors do not warrant a more quantitative regression analysis in our opinion.} $\alpha=0.48\pm0.09$ and $K=-19.1\pm4.4$ (c.f. the H$\beta$ size-luminosity relation of \citealt{ben09} who find $\alpha\simeq 0.52$ and $K\simeq -21.3$). 

To compare the relative sizes of the iron and H$\beta$ emitting-regions, we have remeasured the size of the latter using the prescription employed above, where the H$\beta$ bands follow the definition of \citet[see their table 6]{kas00} and are shown in figures 1-3. The deduced H$\beta$ lags are reported in table 1 and are on average 50\% larger than those determined by \citet[or only 10\% larger if PG\,1700, having the largest lag uncertainties in their work, is discarded]{kas00}. The iron-to-H$\beta$ emission-region size ratio in our sample covers the range $0.9$-$1.5$, i.e., in qualitative agreement with the range of 1.5-1.9 found by \citet[]{bar13} for two low-luminosity sources. Taken together, the results imply an iron emission-region no larger than about twice that of H$\beta$, which is qualitatively consistent with recent theoretical expectations \citep[see their figure 10]{mn12}.

It is further possible to consider the other PGs in the \citet{kas00} sample. The analysis method is identical to that carried out in section 3, but the results are insignificant according to our FR-RSS criterion, as described in section 3 \citep[they are, however, significant according to the algorithm of][]{cz13}. Results for PG\,0052+251, PG\,1226+023, and PG\,0804+761 are shown in figure \ref{rl}. The (K)MCF algorithm could not detect a lagging component in the luminous quasar PG\,1704+608, which is consistent with the relatively small contribution of the iron blends to the spectrum of this source \citep[see their figure 1]{kas00}. Insignificant results were obtained for the fainter PG objects, possibly related to the fact that the sampling period is comparable to the expected lag in those sources \citep[and figure \ref{rl}]{lsst}.

Our deduced size-luminosity relation for the iron blends needs, however, to be regarded with caution: there are only handful of detections, with the results for some sources being potentially affected by sampling (PG\,2130 and PG\,0052+251). Further, while hard to quantify in the present work, biases inherent to the MCF method with respect to standard cross-correlation techniques may be present, although likely at the $\lesssim$20\% level \citep{lsst}, which is consistent with our findings for the H$\beta$ line. To overcome those problems, better sampling is required, for more luminous sources\footnote{The analysis of available datasets for low-luminosity sources is beyond the scope of the present work, and will not alleviate the uncertainties at the high-luminosity end.}.

Our results imply that the iron emitting region is photoionized by the central source also in luminous quasars, and that its size is roughly consistent with that of Balmer lines emission region. This is also in qualitative agreement with the conjecture of  \citet{bg92} based on velocity dispersion considerations. These results, however, are not of sufficient quality to test various scenarios for the origin of the iron-line region \citep{hu08,fer09}. Better spectroscopic data for luminous sources, and/or the use of narrow band filters, combined with photometric reverberation mapping schemes could be very useful for testing the relation found here and arriving at a more coherent picture of the BLR in quasars.

\acknowledgements 
This research has been supported in part by a FP7/IRG PIRG-GA-2009-256434 grant and by grant 927/11 from the Israeli Science Foundation, and the Jack Adler Foundation.  S.~R. is supported at the Technion by the Zeff fellowship.  Research by A.J.B. is supported by NSF grant AST-1108835.

\end{document}